\documentclass[preprint,showpacs,showkeys,amsmath,amssymb,aps,doublespace]{revtex4}
\usepackage[dvips]{graphicx}
\usepackage{setspace}
\usepackage{amsmath}
\usepackage{amsfonts}
\usepackage{amssymb,color}

\usepackage{graphicx,color}
\usepackage{ifpdf}
\usepackage[latin1]{inputenc}

\begin{document}

\title{Phase coexistence far from equilibrium
}

\author{
Ronald Dickman\footnote{email: dickman@fisica.ufmg.br}
}
\address{
Departamento de F\'{\i}sica and
National Institute of Science and Technology for Complex Systems,\\
ICEx, Universidade Federal de Minas Gerais, \\
C. P. 702, 30123-970 Belo Horizonte, Minas Gerais - Brazil
}

\date{\today}

\begin{abstract}

Investigation of simple far-from-equilibrium systems exhibiting phase separation
leads to the conclusion that phase coexistence is not well defined in this context.
This is because the properties of the coexisting nonequilibrium systems depend on how
they are placed in contact, as verified in the driven lattice gas with attractive
interactions, and in the two-temperature lattice gas, under (a) weak global exchange
between uniform systems, and (b) phase-separated (nonuniform) systems.
Thus, far from equilibrium, the notions of universality of phase coexistence (i.e., independence
of how systems exchange particles and/or energy), and of phases with intrinsic properties
(independent of their environment) are lost.
\end{abstract}


\maketitle


\section{Introduction}

Consider a liquid mixture, for example methanol and n-hexane,
cooled below the unmixing temperature, so that two distinct phases emerge,
separated by an interface of microscopic thickness.  The
properties of the coexisting phases can be predicted by equating the chemical potentials
of the two components, in homogeneous samples of each phase.  Similarly, knowing the
chemical potential of the homogeneous liquid and vapor phases as functions of temperature
and pressure permits one to predict the liquid-vapor coexistence curve.
Is the same thermodynamic analysis possible far from equilibrium?

A central issue in nonequilibrium physics
is whether thermodynamics can be extended
to systems far from equilibrium, in particular, to steady states
\cite{oono-paniconi,eyink96,ST,hatano,hayashi,bertin,tomedeoliveira,evans,pradhan,chatterjee}.
A key thermodynamic concept is phase coexistence; indeed, one of the
principal applications of equilibrium thermodynamics is the prediction
of phase coexistence based upon knowledge of the isolated phases.

While there has been much study of nonequilibrium statistical systems and their
associated phase transitions\cite{KLS,zia,marro,odor07,henkel}, the issue of
coexistence between far-from-equilibrium phases remains largely unexplored.
Instances of phase separation in nonequilibrium steady states have been known for
some time, for example in the driven lattice gas with attractive interactions \cite{KLS,zia,marro}.
It appears to have been assumed, implicitly, that nonequilibrium phase coexistence is well defined.

Equilibrium phase coexistence enjoys a high degree of universality.  That is, if phases A and B
with different densities, for example, are known to coexist when present in a single nonuniform system,
(e.g., as a result of phase separation), then they must also coexist if a pair of uniform
systems, prepared in phases A and B, are permitted to exchange particles and/or energy across a fixed
boundary. Here I show, via explicit examples, that which nonequilibrium phases actually coexist depends upon how they
make contact, that is, on how they are permitted to exchange particles.
Thus, far from equilibrium, the universality associated with equilibrium phase coexistence is lost.


Equilibrium statistical mechanics has established a very general notion of ``phase"
in terms of Gibbs measures \cite{georgii}; since such a formalism is not available far from equilibrium,
I revert to simple operational definitions.  I consider systems maintained out of
equilibrium by a steady drive acting on the particles, or via contact with
two reservoirs having different temperatures.
The condition maintaining the
system away from equilibrium is called a {\it drive};  the drive provokes a flux of energy and/or
matter through the system.

A nonequilibrium phase
corresponds to a macroscopic state of a system under a drive, having time-independent, reproducible
properties that vary smoothly with the drive intensity and the external parameters (such as temperature
and chemical potential) associated with the reservoir or reservoirs in contact with the system.
If the macroscopic properties depend in a singular manner on the drive or other
external parameters, the system is said to suffer
a (nonequilibrium) phase transition.

Now consider two systems in nonequilibrium steady states, subject to the same drive and external parameters,
but with distinct, spatially uniform macroscopic properties.  The systems
represent coexisting phases if, when allowed to exchange energy or matter,
the net flux of the quantity or quantities they may
exchange is zero.  Nonequilibrium phase coexistence emerges spontaneously at {\it phase separation},
in which, varying some external parameter, a homogeneous phase becomes unstable, yielding a new stable steady
state containing distinct macroscopic phases separated by a sharp interface.  (The coexisting phases
are clearly free to exchange particles and energy in this situation.)

The stochastic particle system known as the driven lattice gas with attractive interactions
(or Katz-Lebowitz-Spohn (KLS) model \cite{KLS}) provides a simple realization of phase
separation far from equilibrium.  The drive (which favors particle motion along a certain axis),
tends to increase the potential energy; to maintain a steady state, energy flows from the
particle system to a reservoir at a certain temperature $T_R$.  At high temperatures, the
state with uniform density is stable, but (in two or more dimensions) below a certain value of $T_R$, the system
segregates into high- and low-density phases separated by a narrow interface, much as
its equilibrium counterpart (equivalent to the Ising model with fixed total magnetization) undergoes
phase separation.  The high- and low-density regions relax to a state in which their macroscopic properties
are time-independent, despite the presence of the drive.
These ``empirical" facts about the model (known from extensive numerical
simulations) provide an unambiguous example of coexisting nonequilibrium phases.

Suppose a pair of coexisting phases, A and B, are found in a driven, phase-separated system,
and that we prepare uniform systems with the same macroscopic properties (such as density), as
observed in phases A and B, respectively,
subject to the same drive and external parameters as in the phase-separated system.  We may now ask:

1. Are the isolated uniform phases A and B stable?

2. If they are stable, and are allowed to exchange particles and/or energy, will they coexist?

In equilibrium, the stability and coexistence of the uniform phases (given their coexistence
under phase separation), is so ``obvious" that the corresponding questions have hardly been
explored, far from equilibrium.
I investigate these questions in the context of two far-from-equilibrium models: the KLS model mentioned
above, and a two-temperature lattice gas.  Precise definitions of the models are given below.

In the present study, the transition rates for particle exchange between (potentially) coexisting phases are taken as
Sasa-Tasaki (ST) rates \cite{STnote}, which depend on the interaction energy before, but not after, the transition.
I use ST rates for two reasons.  First, as argued in \cite{ST}, these rates correspond to thermally-induced
transitions over an energy barrier: as in the classic Kramers escape problem, the particle cannot ``know"
the energy landscape on the other side of the barrier.  Second, as shown in \cite{incsst}, consistent definitions of
intensive properties (temperature, chemical potential) for the driven system are in general only possible
using ST rates.  (Here ``consistent" is used in the sense of the zeroth law of thermodynamics.)

The balance of this paper is organized as follows.  In section 2 I report on studies of phase coexistence in the KLS
model, followed, in section 3, by studies of the two-temperature lattice gas.  Section 4 presents our conclusions.

\section{KLS model}

The KLS model \cite{KLS,zia,marro}, is a stochastic lattice gas in which each site $i$ of a lattice is either vacant
(occupation variable $\sigma_i = 0$) or occupied ($\sigma_i = 1$).
The interaction energy is:

\begin{equation}
E = - \sum_{\langle i,j \rangle} \sigma_i \sigma_j \;,
\label{EintKLS}
\end{equation}

\noindent where the sum is over nearest-neighbor (NN) pairs of sites; each NN particle pair
lowers the energy by one unit.  In equilibrium, this system is equivalent to the NN ferromagnetic Ising model;
on the square lattice, it exhibits a continuous phase transition at temperature $T_c = T_{c,i}/4 \simeq 0.5673$, where $T_{c,i}$
denotes the Onsager temperature of the Ising model \cite{onsager}.  The KLS model is equipped with a particle-conserving
NN hopping dynamics, and, crucially, a nonequilibrium drive ${\bf D}= D{\bf i}$
imposed via hopping rates favoring
displacements along the $+x$ direction (which must be periodic), and inhibiting those in the opposite sense.
The acceptance probability for a particle displacement $\Delta x$, along the $x$ direction is

\begin{equation}
p_{a,x} = \min\{1, \exp[-\beta(\Delta E - D \Delta x)]\},
\label{pacx}
\end{equation}

\noindent where $\beta = 1/T_R$.
In the present work I study the infinite-$D$ limit: all attempts to hop along the $+x$ direction are accepted
(provided the target site is unoccupied), while hopping in the opposite direction is prohibited.
The acceptance probabilities for hopping by particle $j$ in the transverse directions ($\pm y$) follow the ST prescription:

\begin{equation}
p_{a,y} = \exp[-\beta n_j],
\label{pacy}
\end{equation}

\noindent where $n_j$ is the number of occupied NNs of particle $j$ prior to hopping.   I use ST rates for hopping perpendicular to the
drive because, under phase separation, the interface is along the drive.  (An interface perpendicular to the drive is
unstable \cite{KLS,zia,marro}.)  Thus exchange between coexisting phases is governed by ST rates.

I simulate the KLS model on square lattices of $L_x \times L_y$ sites, with periodic boundaries, using $L=100$, 200 and 400.
(In most of the studies, $L_x = L_y$.)
In the continuous-time stochastic evolution, each particle is equally likely to
be the next to attempt to hop; hopping is always to a NN site.  If the latter is unoccupied, the
particle displacement is accepted with the probabilities defined above.
A Monte Carlo step (MCS) corresponds to one attempted move per lattice site; simulations are run for a total of
1 - $4 \times 10^7$ MCS, with an initial period of 1 - $2 \times 10^7$ MCS for relaxation to the steady state.
Averages are performed over 5 - 10 independent realizations.

A preliminary study revealed that phase separation occurs for temperatures
$T_r < T_c \simeq 0.90$.  (This is somewhat higher than the value, $T_c = 0.769(2)$, for the square lattice
under infinite drive, using Metropolis rates. \cite{marro87}.)  For present purposes a precise result for
$T_c$ is not needed: the only need for an estimate of $T_c$ is to study values well below it,
to avoid finite-size effects large fluctuations associated with the critical region.

\begin{figure}[!htb]
\includegraphics[clip,angle=0,width=0.4\hsize]{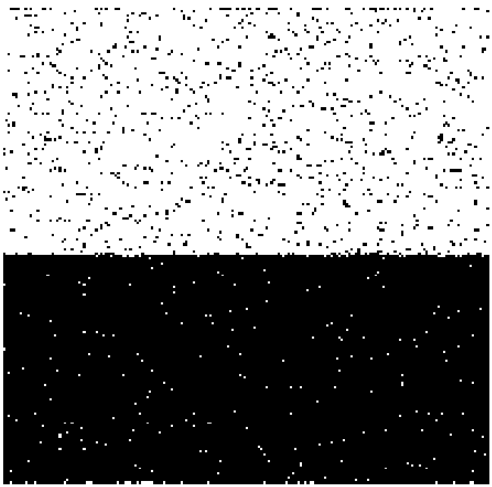}

\caption{\footnotesize{A typical steady-state configuration in the KLS model.
System size $L=200$, temperature $T_R = 0.6$, drive directed to right.
}}
\label{cf800}
\end{figure}

\begin{figure}[!htb]
\includegraphics[clip,angle=0,width=0.8\hsize]{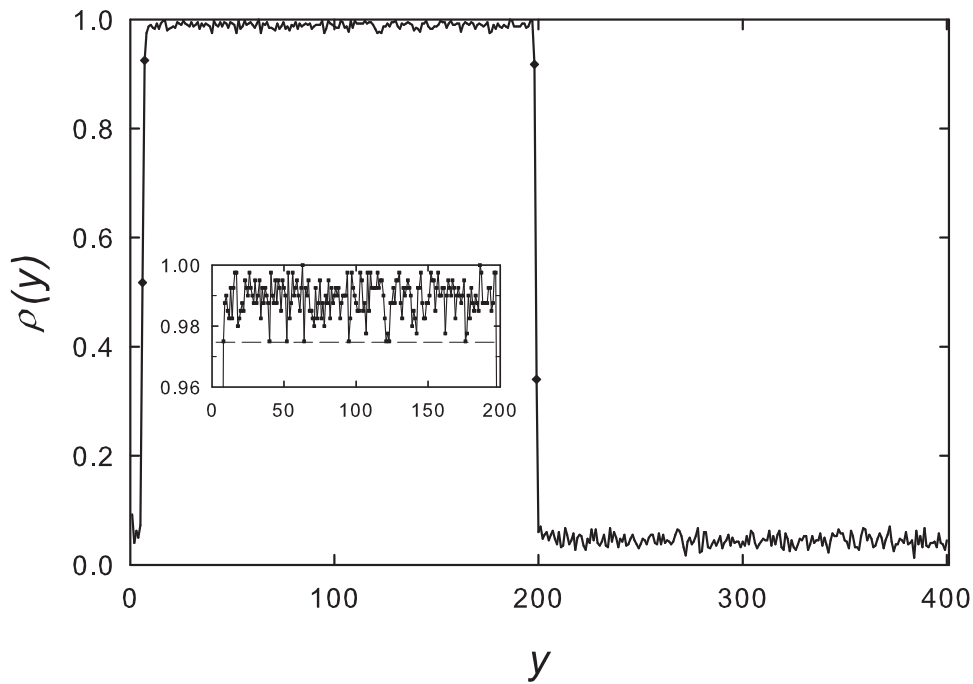}

\caption{\footnotesize{A typical density profile associated with the final configuration in the KLS lattice gas
with Sasa-Tasaki exchange rates.  The profile consists of statistically uniform high- and low-density regions separated by
narrow interfaces (four points).
System size $L=400$, temperature $T_R = 0.6$, total time $3 \times 10^7$ MCS.
Inset: detail of the high-density portion of the profile, showing the criterion used to estimate $\rho_L$, namely,
an average over all points $y_1 \leq y \leq y_2$ such that $\rho(y) \geq \rho_{min}$, the smallest value in the bulk region.
In this case $\rho_{min} = 0.975$, and all points on or above the dashed line are included in the estimate for $\rho_L$,
even thought the leftmost such point might be considered part of the interface.}}
\label{prf807a}
\end{figure}

Starting from a random, statistically uniform distribution of particles, the system evolves to a configuration
a number of dense stripes separated by rarefied regions: the hallmark of phase separation in the KLS model.
To minimize interfacial effects and facilitate determination of the coexisting densities,
I employ a {\it single-stripe} initial configuration in a half-filled system, which relaxes to
a stationary state consisting of a single dense and a single rarefied region (see Fig.~\ref{cf800}).  The coexisting
densities $\rho_L$ and $\rho_V$ are determined through analysis of the final density profile $\rho(y)$ (see Fig.~\ref{prf807a}).
An average over six such final profiles (for $L=400$ and $T_R = 0.6$), taken after a total of $3 \times 10^7$ MCS,
yields $\rho_L = 0.9890(2)$ and $\rho_V = 0.0451(4)$,
where the figures in parentheses denote statistical uncertainties.
(Note that individual profiles are analyzed, not an average profile.)
Profiles taken after only $2 \times 10^7$ MCS
yield densities of 0.9889(2) and 0.0450(5), showing that the densities have indeed relaxed to their stationary
values.  Studies of a smaller system ($L=200$) yield coexisting densities of 0.9893(2) and 0.0459(3), showing that
finite-size effects are minimal.

All profiles exhibit well defined bulk regions, in which the density $\rho(y)$ is free of any significant
linear trend or curvature; the bulk regions are separated by interfaces.  To decide if a given point $y$ (near the
interface) should be taken as part of the bulk, I use the criterion illustrated in the inset of Fig.~\ref{prf807a}:
the minimum value $\rho_{min}$, within the bulk of the high-density region is identified, and the maximal set
$\{y_1,...,y_2\}$ such that $\rho(y) \geq \rho_{min}$ is taken as the bulk, and used to calculate $\rho_L$.  An analogous
criterion (using the maximum density in the low-density region) is employed in calculating $\rho_V$.  Note that
this procedure might be expected to yield a slight underestimate of $\rho_L$, and a slight overestimate of $\rho_V$,
as relatively rare density values, that might arguably be assigned to the interface, are included in the bulk.
The absence of significant finite-size effects nevertheless suggests that such overestimates (or underestimates) are minimal.

In this manner, I determine the coexisting densities for temperatures ranging from 0.5 to 0.7, well below
the critical temperature.
In contrast to the equilibrium lattice gas (or the KLS model using Metropolis
rates), here the coexisting densities do not obey $\rho_L + \rho_V = 1$.  This is because ST rates do not
respect particle-hole symmetry.
I verify that the coexisting densities are insensitive to modest changes in the aspect ratio, that is, using
$L_x = 2L_y$ or $L_x = L_y/2$.

I turn now to studies of {\it composite} systems, consisting of a pair of {\it uniform} systems of the same 
size (with periodic boundaries), at the same reservoir temperature and subject, as
before, to an infinite drive.  The two systems, L and V, are prepared with the densities $\rho_L$ and $\rho_V$
found to coexist in the phase-separated systems.  These systems are allowed to exchange particles with an overall attempt rate $p_r$,
again using ST rates, via weak global exchange.  (It is important to note that there are {\it no interactions} between particles in  
different systems.)
Global exchange means
that any particle in one system may attempt to jump to any site in the other.
Weak exchange corresponds to
the limit $p_r \to 0$.
Important consequences of weak exchange are: (1) the systems in contact are statistically independent;
(2) particle exchange does not provoke spatial inhomogeneities within these systems;
(3) under weak global exchange using ST rates, it is possible to associate a temperature and a chemical potential
with {\it spatially uniform} nonequilibrium steady states in a thermodynamically consistent manner \cite{incsst}.

\begin{figure}[!htb]
\includegraphics[clip,angle=0,width=0.75\hsize]{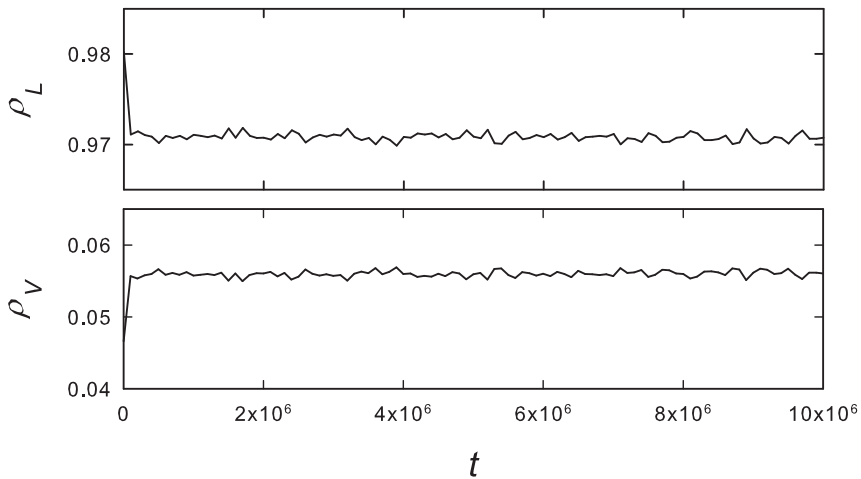}


\caption{\footnotesize{Time evolution of the densities $\rho_L$ and $\rho_V$ in a pair of uniform systems under global
exchange.  Parameters $L=400$, $T_R = 0.55$, $p_r = 0.005$.
}}
\label{rh2sys}
\vspace*{2cm}
\end{figure}

Using the same simulation algorithm as before, I monitor the particle densities in systems L and V over periods of
order $10^7$ MCS; relaxation to stationary values typically requires fewer than $10^6$ MCS.  The final configuration is
checked to assure that both systems remain spatially uniform, i.e., that phase separation has not occurred within either system.
Figure \ref{rh2sys} shows a typical evolution of the densities in the uniform systems under global exchange.
Varying the exchange attempt rate $p_r$ between 0.01 and 0.001, no significant change in the stationary densities
$\rho_L$ and $\rho_V$ is found.

\begin{figure}[!htb]
\includegraphics[clip,angle=0,width=0.75\hsize]{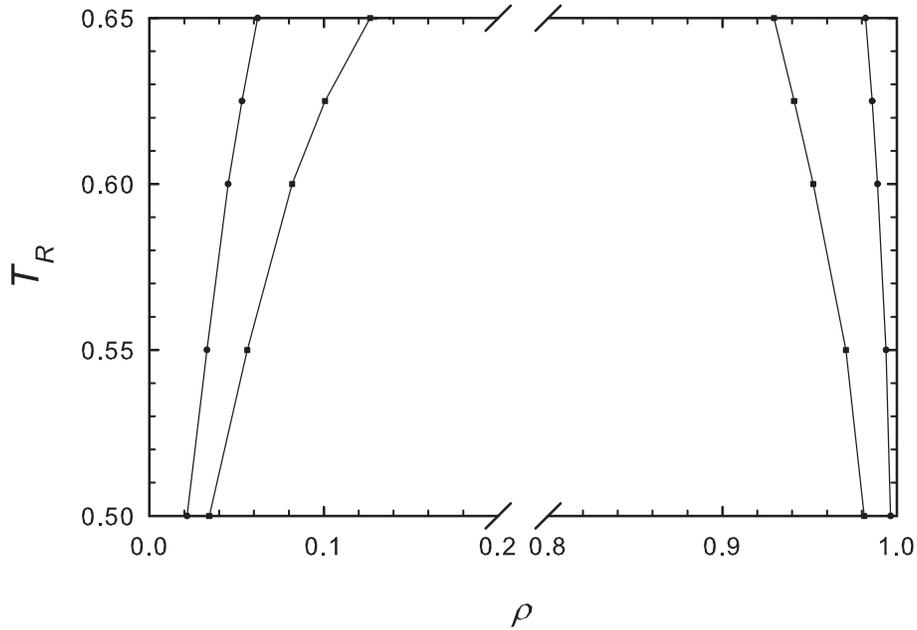}

\caption{\footnotesize{KLS model: coexisting densities $\rho_L$ (right side) and $\rho_V$ (left side) versus
reservoir temperature $T_R$.  The points nearer the vertical axes correspond to the single phase-separated
system; those further from the axes correspond to uniform systems under weak global exchange.  System size $L=400$.
Error bars smaller than symbols.
}}
\label{rhcoex}
\end{figure}

In the stationary state, systems L and V coexist, but {\it not} at the densities
observed under phase separation in a single system.  Compared to the single, phase separated system,
the density $\rho_L$ in the uniform system is consistently smaller, while $\rho_V$ is consistently larger.
(Recall that the procedure for estimating coexisting densities in the single phase-separated system
is likely to {\it underestimate} not overestimate, these differences.)
A quantitative comparison of the coexisting densities is given in Table I, for system size $L=400$ (no significant differences are
observed between studies using $L=200$ and $L=400$);
Fig.~\ref{rhcoex}
illustrates the
general trends.  The coexisting densities under phase separation and under weak global exchange between
uniform systems are clearly incompatible.  While the differences between the coexisting liquid densities $\rho_{L,1}$ and
$\rho_{L,2}$ amount to but a few percent, those for the vapor are considerably larger, with $\rho_{V,2} \simeq 2 \rho_{V,1}$
at the highest temperatures studied.

\begin{figure}[!htb]
\includegraphics[clip,angle=0,width=0.9\hsize]{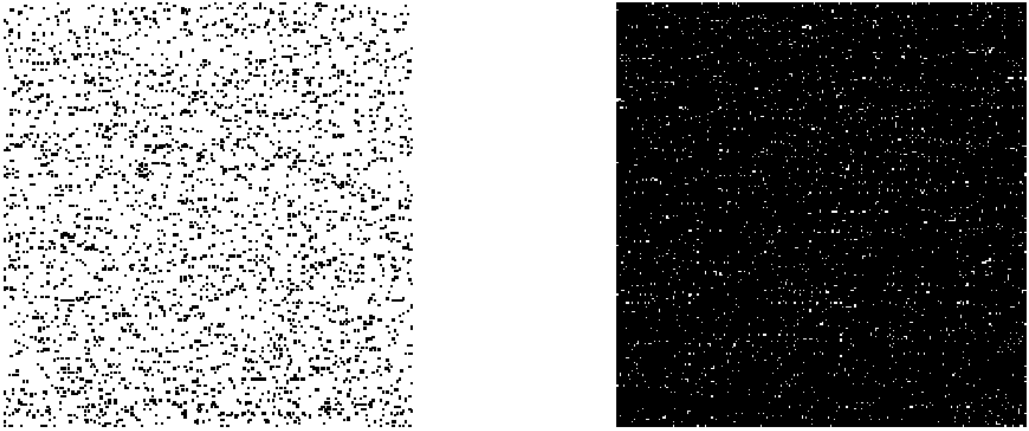}

\caption{\footnotesize{A typical steady-state configuration in the KLS model:
two-system coexistence under weak global exchange (V on left, L on right).
System size $L=200$, temperature $T_R = 0.6$, exchange rate $p_r = 0.003$, drive directed to right.
}}
\label{cfs769}
\end{figure}

\begin{figure}[!htb]
\includegraphics[clip,angle=0,width=0.4\hsize]{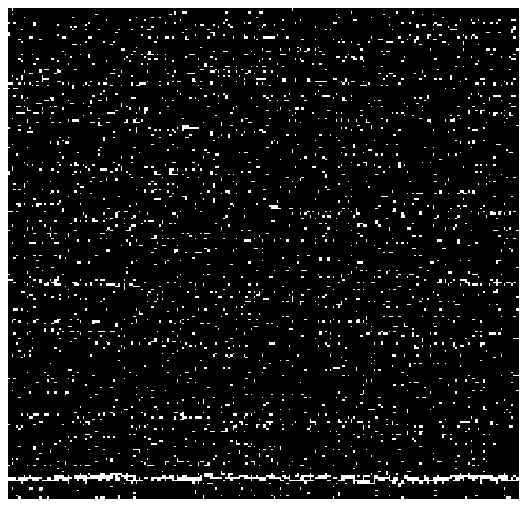}

\caption{\footnotesize{A typical steady-state configuration of the dense system (L) in the KLS model
under weak global exchange.  (The other system (V) is uniform and is not shown.)
Although systems L and V were given initial densities equal to those observed in the single phase-separated
system with the same parameters, an empty strip has formed.
System size $L=200$, temperature $T_R = 0.65$, drive directed to right.
}}
\label{cf921}
\end{figure}

In the two-system studies, the total particle number is determined using the bulk densities observed in the single,
phase-separated system: $N = N_L + N_V = 2L^2(\rho_{L,1} + \rho_{V,1})$.
(Initially, $N_L = L^2$.)
During the stochastic evolution
under exchange, the particle numbers $N_L$ and $N_V$ quickly attain stationary values, subject, of course,
to the constraint of fixed $N$.  For temperatures $T_R \leq 0.6$, the final configurations are spatially uniform
(see Fig.~\ref{cfs769}).
At higher temperatures, however, one observes the formation of stripes: one or both systems have undergone phase
separation, showing that the phase that is stable in the single system is {\it unstable} in the composite system (see Fig.~\ref{cf921}).
To determine the coexisting densities in this case, I search for values of $N$ such that both systems remain
uniform.  In Table I, only the values for $T_R \geq 0.625$ were obtained in this manner; the densities listed are the
smallest for which the composite system is stable.  Stable coexistence in the composite system is not observed
for {\it any} density, for $T_R = 0.7$.
(The situation is reminiscent of a somewhat different system, studied by Achahbar and coauthors, \cite{achahbar95,achahbar96},
who employed exchange between pairs of corresponding sites, rather than global exchange, in a pair of KLS models.  These authors
observed phase separation within each system (in corresponding regions), for temperatures between the critical values for the driven and
undriven systems, and phase coexistence between uniform systems at temperatures below the critical value for the
undriven model.)

\begin{table}[h]
\begin{center}
\begin{tabular}{|c|c|c|c|c|} \hline
$T_R$ & $\rho_{L,1}$ & $\rho_{V,1}$ & $\rho_{L,2}$ & $\rho_{V,2}$ \\ \hline\hline
 0.50  & 0.9965(1)    &   0.0216(1)  &  0.9826(1)   & 0.0360(1)   \\ \hline
 0.55  & 0.9939(1)    &   0.0330(2)  &  0.97088(2)  & 0.05603(2)  \\ \hline
 0.60  & 0.9890(2)    &   0.0451(4)  &  0.9521(1)   & 0.0818(1)   \\ \hline
 0.625 & 0.9859(1)    &   0.0531(1)  &  0.9411(1)   & 0.1008(1)   \\ \hline
 0.65  & 0.9821(2)    &   0.0621(1)  &  0.9297(1)   & 0.1266(1)  \\
\hline \hline
\end{tabular}
\end{center}
\label{tb1}
\noindent
\caption{KLS model: coexisting densities in a single phase-separated system ($\rho_{L,1}$, $\rho_{V,1})$,
and in a pair of uniform systems under weak global exchange ($\rho_{L,2}$, $\rho_{V,2})$.}
\end{table}

\begin{figure}[!htb]
\includegraphics[clip,angle=0,width=0.6\hsize]{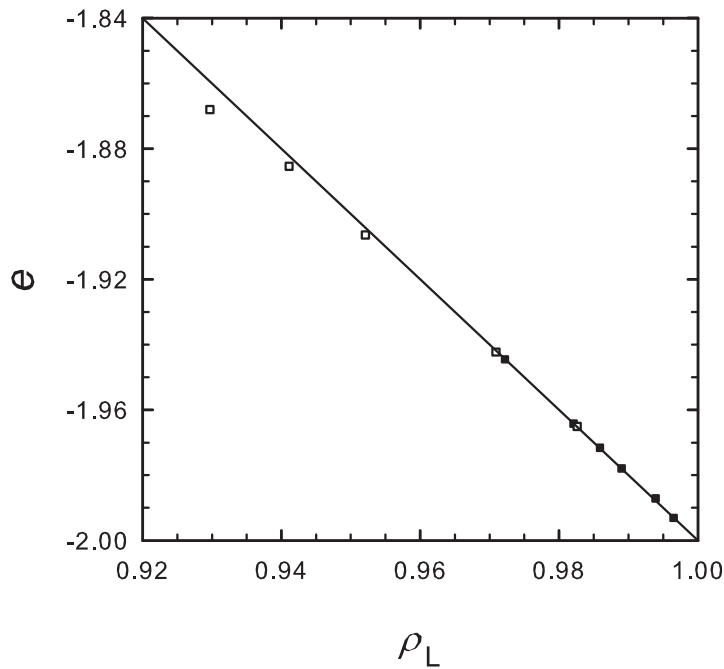}

\caption{\footnotesize{KLS model: mean interaction energy per particle $e$ versus density $\rho$ at coexistence
in the high-density phase, in the single system (filled symbols) and the composite system (open symbols).  Solid line: random-mixing
prediction, $e = 2\rho$.
}}
\label{evrl}
\end{figure}

\begin{figure}[!htb]
\includegraphics[clip,angle=0,width=0.6\hsize]{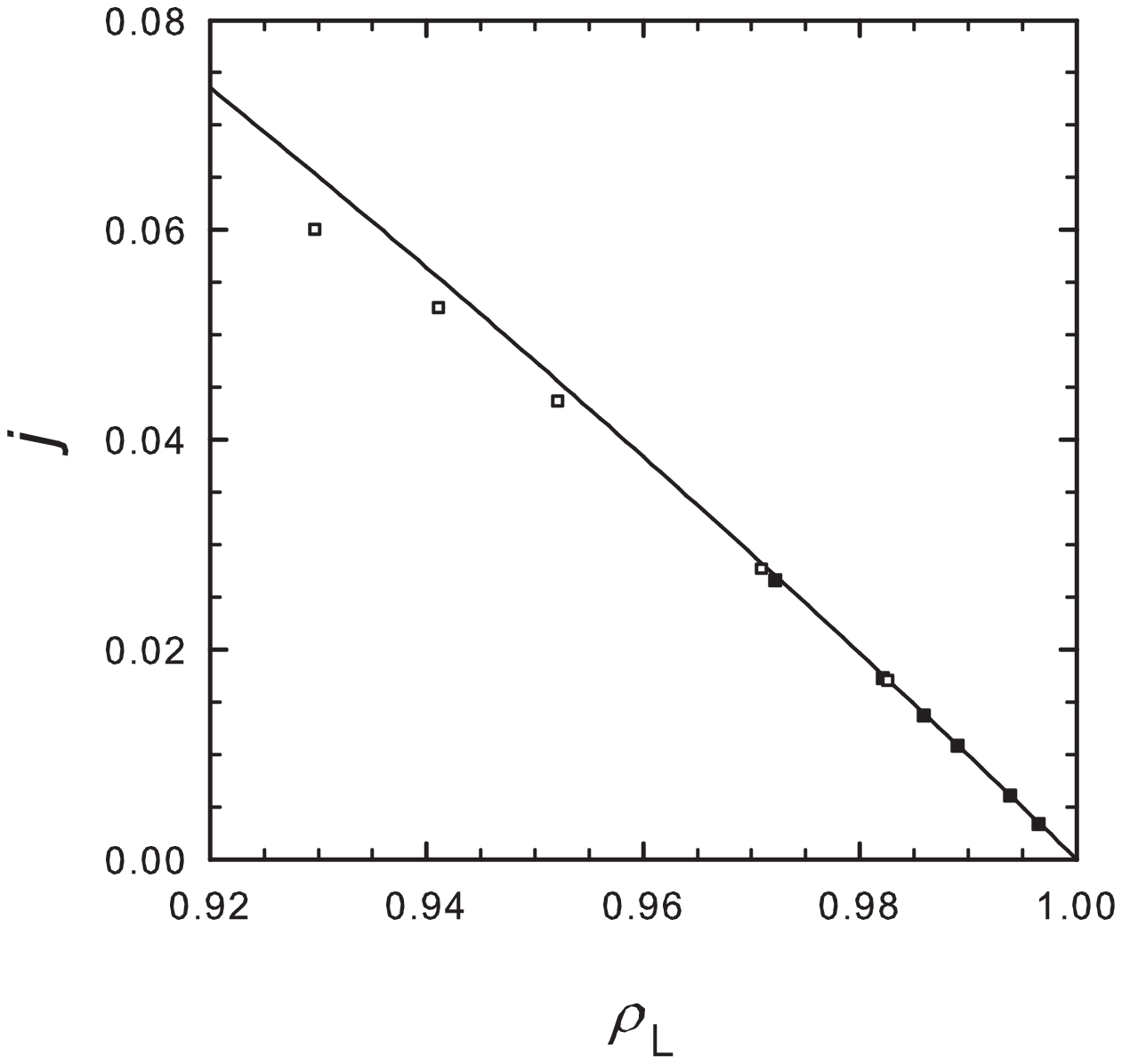}

\caption{\footnotesize{KLS model: mean current $j$ versus density $\rho$ at coexistence
in the high-density phase, in the single system (filled symbols) and the composite system (open symbols).  Solid line: random-mixing
prediction, $j = \rho(1-\rho)$.
}}
\label{jvrl}
\end{figure}

\begin{figure}[!htb]
\includegraphics[clip,angle=0,width=0.6\hsize]{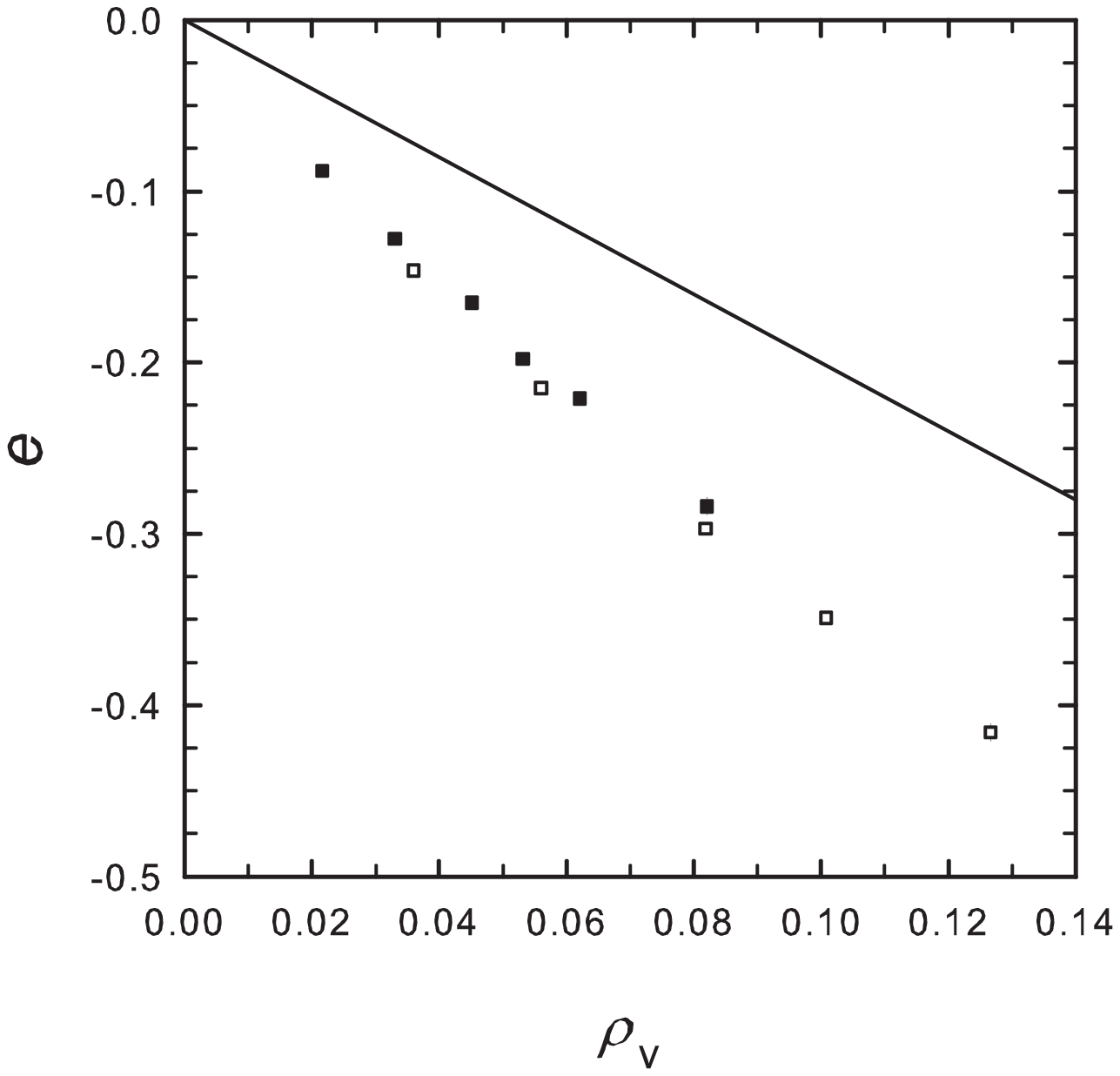}

\caption{\footnotesize{KLS model: mean interaction energy per particle $e$ versus density $\rho$ at coexistence
in the low-density phase, in the single system (filled symbols) and the composite system (open symbols).  Solid line: random-mixing
prediction, $e = 2\rho$.
}}
\label{evrv}
\end{figure}

\begin{figure}[!htb]
\includegraphics[clip,angle=0,width=0.6\hsize]{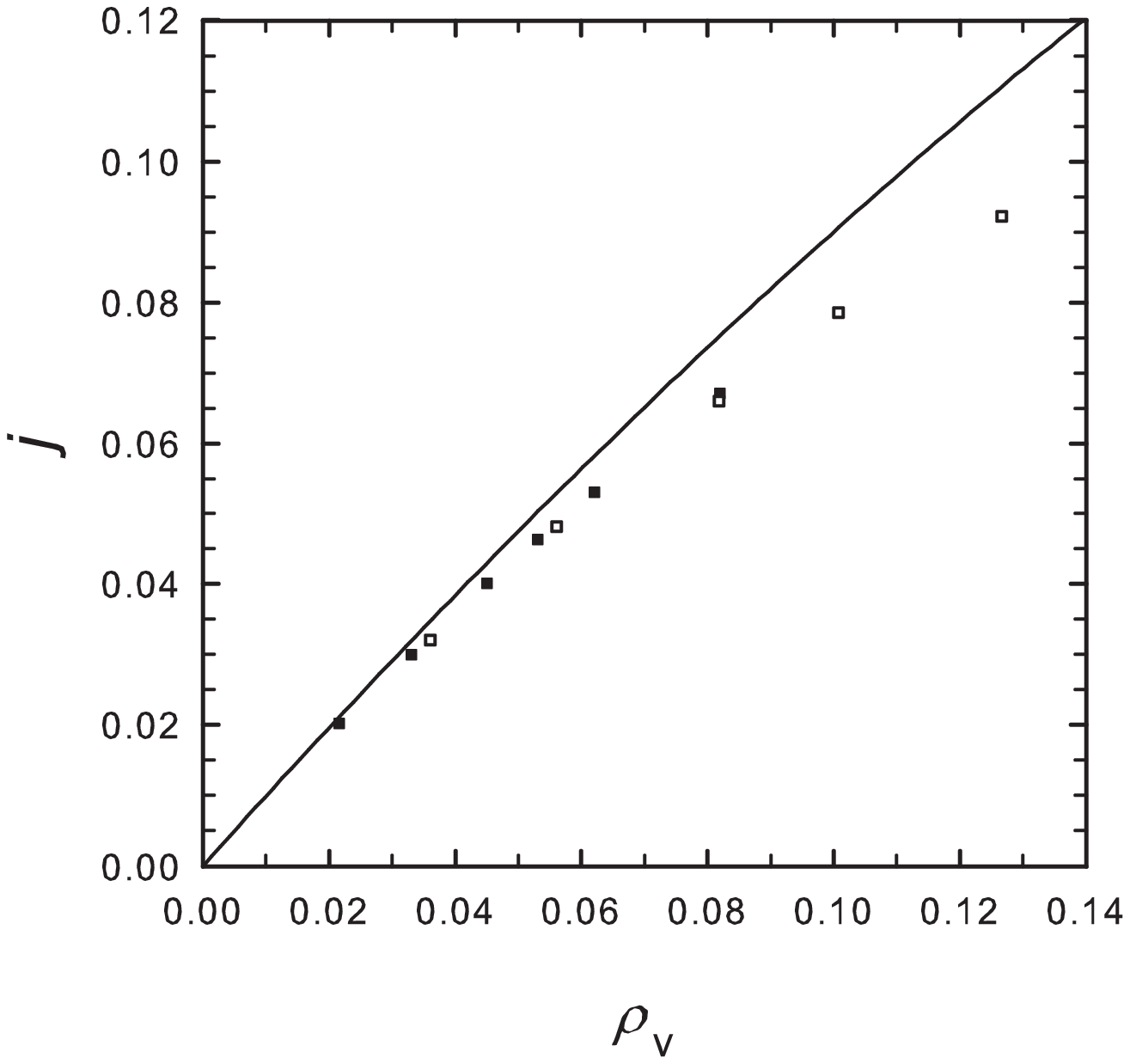}

\caption{\footnotesize{KLS model: mean current $j$ versus density $\rho$ at coexistence
in the low-density phase, in the single system (filled symbols) and the composite system (open symbols).  Solid line: random-mixing
prediction, $j = \rho(1-\rho)$.
}}
\label{jvrv}
\end{figure}

Discrepancies between coexisting densities in the single and composite systems naturally lead to
differences in other stationary macroscopic properties, such as the mean interaction energy per particle $e$, and mean current density $j$.
Plotting these functions versus particle density $\rho$ affords some insight into the nature of the coexisting phases.  In the high-density
(L) phase, both $e$ and $j$ appear to be functions of $\rho$ alone, and are well-approximated by the random-mixing
expressions $e = 2\rho$ and $j = \rho(1-\rho)$, as shown in Figs.~\ref{evrl} and \ref{jvrl}.
The corresponding plots for the low-density (V) phase (see Figs.~\ref{evrv} and \ref{jvrv}) show substantial
differences between the single- and composite-system
properties, and strong deviations from the random-mixing predictions.  This is rather natural, since we expect
$e$ and $j$ to depend on density {\it and} temperature, and
when the densities are equal, the temperatures are different.  The insensitivity to temperature evident
in the L phase is consistent with its approximation to a random mixture, which would appear to be
a feature of a density close to unity and a correspondingly short correlation length.

Summarizing results for the KLS model, at all temperatures studied, there is a clear discrepancy between the
densities characterizing coexisting phase in a single, phase-separated system, and those associated with a pair of
uniform systems that coexist under global exchange.

\section{Two-temperature lattice gas}

The KLS model is strongly anisotropic, and features a particle current along one of the lattice axes.  It is natural to
ask whether these features are somehow responsible for the nonuniversality of phase coexistence documented above.
This question motivates study of a second far-from-equilibrium system, a two-temperature lattice gas (TTLG).
The interaction energy is again given by Eq.~\ref{EintKLS}.  The stochastic evolution is via NN particle
hopping, with acceptance probabilities as in Eq.~\ref{pacy}, in {\it both} the $x$ and $y$ directions.
The nonequilibrium drive in this case takes the form of two reservoir temperatures, $T_A$ and $T_B$, associated
with sublattices A and B.  (Sublattices A and B comprise the sets of sites $(i,j)$ with $i+j$ even and odd,
respectively.)  The case $T_A = T_B$ corresponds to the equilibrium lattice gas, equivalent to the NN Ising model.
For $T_A \neq T_B$ the system cannot reach equilibrium.  Note however that the system is isotropic and that there is no
net particle current.  (The drive $T_A - T_B$ induces an energy flux between the sublattices.)

The TTLG exhibits a line of Isinglike phase transitions in the $T_A - T_B$ plane \cite{blote90,heringa94,bakaev95}.
To observe phase separation in the form of coexisting strips, I initialize the system with all sites in
half of the system ($j \leq L/2$) occupied, and the other half vacant.
Coexisting densities are determined from the final density profile, as in the KLS studies.  I verify that
the estimates for system sizes of $L=400$ and $L=200$ agree to within uncertainty.  Next, a pair of uniform systems,
with particle densities corresponding to the coexisting densities found under phase separation, are prepared,
and studied in simulations of 1 - 3 $\times 10^7$ MCS.  As before, the particle densities in the coexisting
systems are monitored; final configurations are checked for uniformity.

\begin{figure}[!htb]
\includegraphics[clip,angle=0,width=0.4\hsize]{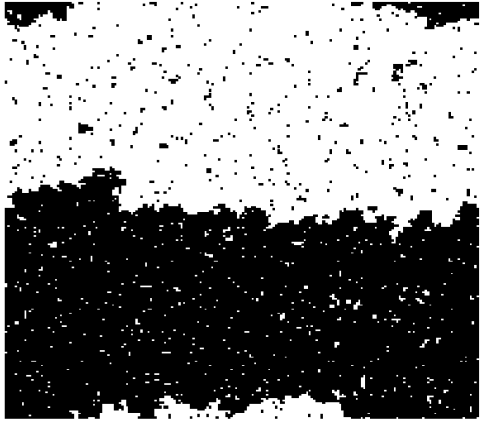}

\caption{\footnotesize{A typical steady-state configuration in the TTLG.
System size $L=200$, temperatures $T_A = 0.4$, $T_B = 0.55$.
}}
\label{cf111}
\end{figure}

\begin{figure}[!htb]
\includegraphics[clip,angle=0,width=0.8\hsize]{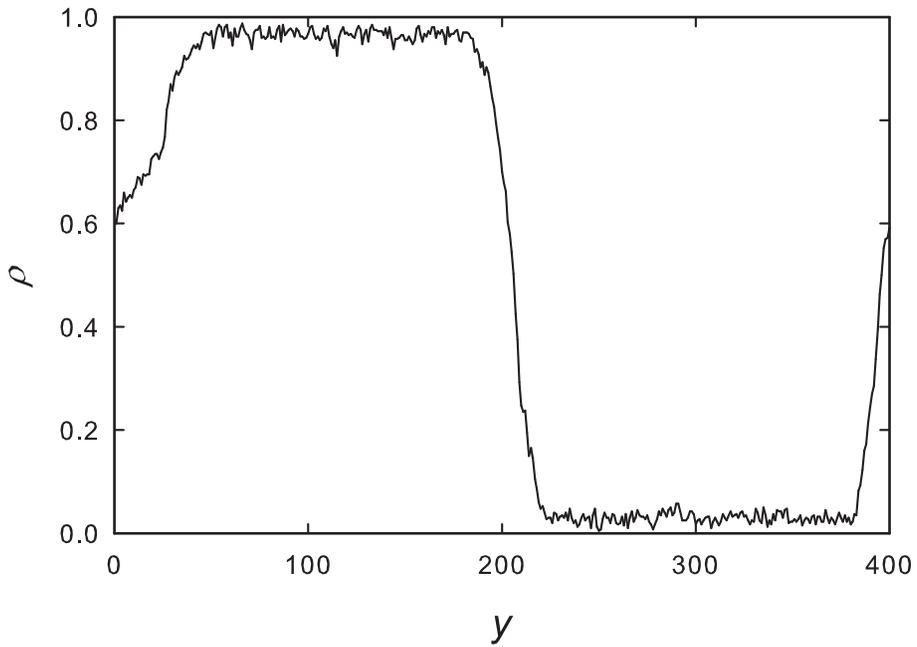}

\caption{\footnotesize{A typical density profile associated with the final configuration in the TTLG.
System size $L=400$, $T_A = 0.4$, $T_B = 0.55$, total time $6 \times 10^7$ MCS.
}}
\label{prf131}
\end{figure}

As a test, I set $T_A = T_B = 0.5$, and observe $\rho_{L,1} = 0.9552(4)$ and $\rho_{V,1} = 0.04454(2)$ in the single, 
phase-separated system, whilst
coexistence in the composite system yields $\rho_{L,2} = 0.9556(3)$ and $\rho_{V,2} = 0.0444(3)$.  Thus, as expected,
at equilibrium the two modes of coexistence yield phases with densities that agree to within uncertainty,
and, moreover, $\rho_L + \rho_V = 1$.  For unequal sublattice temperatures there are discrepancies.  Studies using
$T_A = 0.4$ and $T_B = 0.55$, for example, yield $\rho_{L,1} = 0.9644(3)$ and $\rho_{V,1} = 0.03001(3)$, in a single phase-separated
system.  A typical configuration is shown in Fig.~\ref{cf111}.  Although the interface in now rough (since there is no particle
current), well defined bulk phases are present.  The coexisting densities are again obtained via analysis of the final
density profiles; a typical profile is shown in Fig.~\ref{prf131}.
As before, coexisting densities between two uniform systems under weak global exchange are determined by
monitoring the particle numbers in the respective systems, once they have attained a steady state.
For these temperatures, the composite-system studies
furnish $\rho_{L,2} = 0.9511(15)$ and $\rho_{V,2} = 0.0333(4)$, clearly incompatible with the values
of $\rho_{L,1}$ and $\rho_{V,1}$ cited above.  Similar inconsistencies are found at other values of $T_A$ and $T_B > T_A$,
as shown in Table II.

In the TTLG with sublattice temperatures $T_A = 0.4$ and $T_B = 0.6$, separation into well defined coexisting phases
in a single system is observed, but it appears to be impossible to stabilize coexistence of two uniform
phases in a composite system, regardless of how the total density is varied.  This may be related to the fact that
$T_B$ is above the critical temperature of the equilibrium lattice gas.

In both the KLS and TTLG systems, the difference (at a given temperature) between coexisting densities
is smaller in the composite system than in the single system.  The reason for this tendency, while not obvious,
is presumably connected with the presence of an interface in the single system and its absence in  the composite system.
One might speculate that the one-step transitions taking a particle directly from one bulk phase to the other facilitate
exchange in the composite system, as compared with the series of ``uphill" displacements required to transport particles across the interface
in the single system.  This line of argument appears to fail, however, when we note that for very small exchange rates $p_r$,
the rate of transfers between phases in the composite system is much smaller than that in the single system.
The coexisting densities in the composite system are nevertheless essentially independent of $p_r$ as it tends to zero.

\begin{table}[h]
\begin{center}
\begin{tabular}{|c|c|c|c|c|c|} \hline
$T_A$ & $T_B$ & $\rho_{L,1}$ & $\rho_{V,1}$ & $\rho_{L,2}$ & $\rho_{V,2}$ \\ \hline\hline
 0.30 & 0.50  & 0.9828(2)    &  0.0098(3)   &  0.9709(1)   & 0.01202(1)   \\ \hline
 0.35 & 0.55  & 0.9706(1)    &  0.02053(2)  &  0.951(3)    & 0.0240(16)   \\ \hline
 0.40 & 0.55  & 0.9644(3)    &  0.03001(3)  &  0.9511(15)  & 0.0333(4)    \\
\hline \hline
\end{tabular}
\end{center}
\label{tb2}
\noindent
\caption{TTLG: coexisting densities in a single phase-separated system ($\rho_{L,1}$, $\rho_{V,1})$,
and in a pair of uniform systems under weak global exchange ($\rho_{L,2}$, $\rho_{V,2})$.}
\end{table}

\section{Conclusions}

Studies of two of the simplest nonequilibrium models exhibiting phase separation are found to have coexisting
bulk properties that depend on how the phases coexist.
Violations of universality in coexistence are observed in two systems, one anisotropic and bearing a particle current,
the other isotropic and free of such a current, suggesting that such violations are generic to phase coexistence far from equilibrium.
Put another way, the results suggest that the notion of {\it phase} as a state of matter with bulk properties
depending only on a small set of intensive parameters does not apply far from equilibrium.  The properties of
the coexsting phases in the KLS and TTLG models depend not only upon the reservoir temperature(s) and the drive strength, but
on the spatial relation between the phases.  Of course, many further possibilities exist, each
presumably leading to a different set of coexisting densities.  For example, allowing long-range hopping in the
single phase-separated system destabilizes the coexisting regions, leading to multiple-stripe configurations.

As shown in previous work on the KLS model and the lattice gas with nearest-neighbor exclusion (NNE) \cite{incsst,SST2},
it is possible to implement steady-state thermodynamics (SST) in a consistent manner in spatially uniform systems, if (and in general only if)
ST exchange rates are used, and provided these rates tend to zero.  By contrast, SST is inconsistent and without
predictive value in spatially nonuniform NNE models.  Here, rather than imposing nonuniformities via a nonuniform
drive, or walls, they arise spontaneously due to phase separation.  Our results imply that in this case as well,
intensive properties (whose equality would predict phase coexistence) cannot be assigned to the coexisting
phases consistently.  While one could in principle assign a temperature and chemical potential to spatially
uniform systems via coexistence with thermal and particle reservoirs \cite{incsst}, these parameters would be of no use in predicting
phase coexistence in a single, phase-separated system.
Our results
suggest a new viewpoint regarding nonequilibrium phases.  From the equilibrium context, we are used to thinking of a phase
as having intrinsic properties, immutable under coexistence with another phase.
Far from equilibrium, the properties
of coexisting phases depend not only on the control parameters (drive, reservoir temperature), but also 
on precisely how exchanges of matter and energy between the phases are realized.


\newpage
\noindent {\bf Acknowledgments}

I thank Royce K. P. Zia for helpful comments.
This work was supported by CNPq and CAPES, Brazil.
\vspace{2em}



\begin{thebibliography}{100}


\bibitem{oono-paniconi}
        Y. Oono and M. Paniconi,
        Prog. Th. Phys. Supp. {\bf 130}, 29 (1998).

\bibitem{hatano}
        T. Hatano and S. Sasa,
        Phys. Rev. Lett. {\bf 86}, 3463 (2001).

\bibitem{hayashi}
        K. Hayashi and S. Sasa,
        Phys. Rev. E{\bf 68}, 035104 (2003).

\bibitem{bertin}
        E. Bertin, K. Martens, O. Dauchot, and M. Droz,
        Phys. Rev. E{\bf 75}, 031120 (2007).

\bibitem{tomedeoliveira}
        T. Tom\'e and M.~J. de Oliveira,
        Phys. Rev. Lett. {\bf 108}, 020601 (2012).


\bibitem{evans}
        S.~R. Williams, D.~J. Searles, and D.~J. Evans,
        Mol. Sim. {\bf 40}, 208 (2014).

\bibitem{pradhan}
        P. Pradhan, R. Ramsperger, and U. Seifert,
        Phys. Rev. E{\bf 84}, 041104 (2011).

\bibitem{chatterjee}
        S. Chatterjee, P. Pradhan, and P.~K. Mohanty,
        Phys. Rev. Lett. {\bf 102}, 030601 (2014).

\bibitem{eyink96}
        G.~L. Eyink, J.~L. Lebowitz, and H. Spohn,
        J. Stat. Phys. {\bf 83}, 285 (1996).

\bibitem{ST}
        S. Sasa and H. Tasaki,
        J. Stat. Phys. {\bf 125}, 125 (2006).

\bibitem{KLS}
        S. Katz, J.~L. Lebowitz, and H. Spohn,
        Phys. Rev. B{\bf 28}, 1655 (1983);
        J. Stat. Phys. {\bf 34}, 497 (1984).

\bibitem{zia}
        B. Schmittmann and R.~K.~P. Zia,
        {\it Statistical Mechanics of Driven Diffusive Systems},
        Vol. 17 of {\it Phase Transitions and Critical Phenomena},
        edited by C. Domb and J. L. Lebowitz
        (Academic Press, London, 1995).

\bibitem{marro}
        J. Marro and R. Dickman,
        {\it Nonequilibrium Phase Transitions in Lattice Models}
        (Cambridge University Press, Cambridge, 1999).

\bibitem{odor07}
        G. \'Odor,
        {\it Universality In Nonequilibrium Lattice Systems: Theoretical Foundations}
        (World Scientific,Singapore, 2007)

\bibitem{henkel}
        M. Henkel, H. Hinrichsen and S. Lubeck,
        {\it Non-Equilibrium Phase Transitions Volume I: Absorbing Phase Transitions}
        (Springer-Verlag, The Netherlands, 2008).


\bibitem{georgii}
        H.-O. Georgii,
        {\it Gibbs measures and phase transitions}, 2nd ed.,
        (Walter de Gruyter, Berlin, 2011).

\bibitem{STnote}
        The Sasa-Tasaki (ST) rate for transfer of a particle from system A to system B
        depends exclusively on parameters associated with $A$, and vice-versa \cite{ST}.
        Imposing detailed balance, the ST rate for this transition takes the form
        $w_{ST} = \epsilon  \exp [\beta (\mu_A - E_A)]$, where $E_A$ is the
        energy of interaction between the particle and its neighbors, and $\epsilon$
        is an arbitrary rate factor.  The ST expression can be seen as resulting from a
        high energy barrier between $A$ and $B$; the particle first makes a transition from $A$
        to the barrier, and from there to $B$.

\bibitem{incsst}
        R. Dickman and R. Motai,
        Phys. Rev. E {\bf 89}, 032134 (2014).

\bibitem{onsager}
        L. Onsager,
        Phys. Rev. {\bf 37}, 405 (1931); {\bf 38}, 2265 (1931).


\bibitem{marro87}
        J. Marro, J. L. Vall\'es, and J. M. Gonzalez-Miranda,
        Phys. Rev. B{\bf 35}, 3372 (1987).

\bibitem{achahbar95}
        A. Achahbar and J. Marro,
        J. Stat. Phys. {\bf 78}, 1493 (1995).

\bibitem{achahbar96}
        A. Achahbar, P.~L. Garrido, and J. Marro,
        Molec. Phys. {\bf 88}, 1157 (1996).

\bibitem{blote90}
        H.~W.~J. Blote, J.~R. Heringa, A. Hoogland, and R.~K.~P. Zia,
        J. Phys. A {\bf 23} 3799 (1990).

\bibitem{heringa94}
        J.~R. Heringa, H.~W.~J. Bl\"ote, and A. Hoogland,
        Int. J. Mod. Phys. C {\bf 05}, 589 (1994).

\bibitem{bakaev95}
        A.~V. Bakaev and V.~I. Kabanovich,
        Int. J. Mod. Phys. B {\bf 09}, 3181 (1995).

\bibitem{SST2}
        R. Dickman,
        Phys. Rev. E {\bf 90}, 062123 (2014).




















\end{thebibliography}
\end{document}